# Quantum-Resistant Cryptography

John Preuß Mattsson, Ben Smeets and Erik Thormarker[1]

Ericsson Security Research

**Abstract.** Quantum-resistant cryptography is cryptography that aims to deliver cryptographic functions and protocols that remain secure even if large-scale fault-tolerant quantum computers are built. NIST will soon announce the first selected public-key cryptography algorithms in its Post-Quantum Cryptography (PQC) standardization which is the most important current effort in the field of quantum-resistant cryptography. This report provides an overview to security experts who do not yet have a deep understanding of quantum-resistant cryptography. It surveys the computational model of quantum computers; the quantum algorithms that affect cryptography the most; the risk of Cryptographically Relevant Quantum Computers (CRQCs) being built; the security of symmetric and public-key cryptography in the presence of CRQCs; the NIST PQC standardization effort; the migration to quantum-resistant public-key cryptography; the relevance of Quantum Key Distribution as a complement to conventional cryptography; and the relevance of Quantum Random Number Generators as a complement to current hardware Random Number Generators.

## Contents

---

[1] erik.thormarker@ericsson.com

# 1    Scope

Quantum-resistant cryptography is cryptography that aims to deliver cryptographic functions and protocols that remain secure even if large-scale fault-tolerant quantum computers are built. In this report we summarize the current state of quantum-resistant cryptography and report on the progress of the most important effort in this area: the NIST Post-Quantum Cryptography standardization. We also give a summary of the security and practicality of Quantum Key Distribution (QKD) since it has been mentioned as a hypothetical quantum-resistant solution in the past. As part of the background for the report we give a high-level overview of the computational model for quantum computers, the quantum algorithms that affect cryptography the most, and what we know about the progress of building machines that can execute these algorithms. This report is not an implementation guide, rather it aims to provide an overview of the state-of-the-art and what we can expect in the coming years.

A paragraph or section marked with 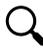 contains information that is slightly more technical than other parts of the report. These parts are not necessary to read to understand the overall report.

We include a discussion on quantum random number generators (QRNGs) in Appendix A. This is a technology that is sometimes mentioned in connection with QKD and quantum computers, however QRNGs do not really fit into our overall scope here since our current hardware randomness generator technology is secure and quantum-resistant. Also, QRNGs themselves do not solve the current urgent question to have secure random number generators in virtualized (VMs, containerized) products. However, as we discuss in Appendix A, if trustworthy vendors make QRNG technology in the future that is as well-understood and certified to the same degree as common RNG technology, then QRNGs could be evaluated as alternatives to common RNG technology.

## 1.1    Terminology

We call our current traditional computers "classical". A "classical attacker" is one that uses a classical computer and not a quantum one. Similarly, a "classical algorithm" is one that can execute on a classical computer.

# 2 Summary

- A Cryptographically Relevant Quantum Computer (CRQC) is a quantum computer of sufficient size and fault tolerance to break today's public-key cryptography using Shor's algorithm. Quantum-resistant cryptography is cryptography that is believed to remain secure even if CRQCs are built.

- The risk that CRQCs will be built means that currently deployed public-key cryptography needs to be replaced with quantum-resistant alternatives. For example, information encrypted using today's public-key cryptography can be recorded by attackers and later attacked if QRQCs can be built. The potential damage CRQCs may inflict here is the core of the motivation to seek for counter measures even if we have uncertainties around when and if these computers can be built. Updating deployed systems that are using public-key cryptography can also take many years.

- It is very unclear when, or even if, a CRQC will ever be built. The gap between today's quantum computers and envisioned CRQCs is huge, and the field of building quantum computers faces some near-term challenges such as for example no known applications for the Noisy Intermediate-Scale Quantum (NISQ) computers that are expected to be built these coming years.

- There is a consensus in the security community that Post-Quantum Cryptography (PQC) as is being standardized by NIST is the best quantum-resistant solution to replace today's public-key cryptography.

- The best general strategy now is to follow the NIST PQC standardization and see how the progress and outcome is received by the security community and other standardization organizations. Industry products can also consider information from the public-key cryptography migration projects that are launching, such as the one by NIST [68].

- Each one of the new public-key algorithms in the NIST PQC Standardization have some property that hinders it from being drop-in-replacement for today's public-key cryptography. The new algorithms for example have larger communication overhead than elliptic curve cryptography, and many of them also have larger communication overhead than RSA.

- The security of symmetric cryptography (including cryptographic hash functions) is essentially unaffected by CRQCs (including key sizes). While Shor's algorithm is expected to break today's public-key cryptography in a matter of hours on a single CRQC, the algorithm that applies to symmetric cryptography, Grover's algorithm, is expected to have a hypothetical running time of many billions of years on a similarly sized CRQC.

- There is a consensus in the security community that QKD has many fundamental issues that would need to be solved before being considered as a secure complement to conventional cryptography.

- Quantum random number generators solve no real issue with our current hardware random number generators (RNGs). If trustworthy vendors make QRNG technology in the future that is as well-understood and certified to the same degree as common RNG technology, then QRNGs could be evaluated as alternatives to common RNG technology.

# 3      Quantum computers

Following [131], we define a Cryptographically Relevant Quantum Computer (CRQC) as a quantum computer of sufficient size and fault tolerance to break today's public-key cryptography using Shor's quantum algorithm[2].

## 3.1      The computational model of quantum computers

> ➢ Quantum computers are not general-purpose (super) computers, rather they are potential special-purpose machines for certain problems where we can leverage their peculiar nature through clever quantum algorithms.

🔍 This section contains an informal high-level introduction to the quantum computational model. It is not necessary to read to understand material outside of this section or the conclusions in the subsequent subsections of this section. It is however necessary to read the section in order to understand a few of the details in those other subsections.

Quantum algorithms are typically expressed in the circuit model. In this model, one typically starts from a set of qubits, applies a series of quantum gates to them, and measures in the end to produce an output. The state of a qubit $Q$ is a linear combination of the states $\overline{0}$ and $\overline{1}$ with coefficients $\alpha$ and $\beta$ that are complex numbers[3]. That is, $Q = \alpha\overline{0} + \beta\overline{1}$. The states $\overline{0}$ and $\overline{1}$ can be thought of as the 0 and 1 state of a classical bit. It is sometimes said that the state of the qubit is "between" $\overline{0}$ and $\overline{1}$. Figure 1 shows some example operations in a very simple circuit.

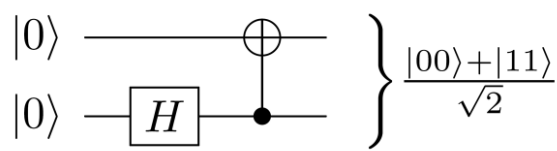

Figure 1: Example operations in the circuit model.

---

[2] One could also define a CRQC as a quantum computer that is sufficiently powerful to threaten today's cryptography in general, thereby also covering, for example, implementations of Grover's algorithm against symmetric cryptography. However, the importance of Shor's algorithm (see Section 3.2) together with the extremely long running time of Grover's algorithm (see Section 3.3), for cryptographically relevant problems, makes it reasonable to let the purpose of the definition be a kind of shorthand to describe a machine that can threaten today's public key cryptography using Shor's algorithm. A special case that would be somewhat problematic with this definition would be if quantum computers turned out to be much faster than classical computers and if the possible circuit width (see Section 3.1) of implementations turned out to be restricted in a way so that cryptographically relevant implementations of Grover's algorithm that target a low width usage (see e.g., [132]) are possible, while cryptographically relevant implementations of Shor's algorithm are not. It is implicitly assumed in [131] that CRQCs can execute Shor's algorithm on cryptographically relevant problems. We make this explicit in our definition.

[3] Mathematically speaking, Q is a unit vector in a two-dimensional complex vector space.

At the start of the circuit (to the left in the figure), we have two qubits, and each qubit is independently in the simple state $\bar{0}$. That is, $Q = \alpha\bar{0} + \beta\bar{1}$, with $\alpha = 1$ and $\beta = 0$ in the notation from above. As we apply quantum gates, the qubits become <u>entangled</u>, which roughly means that their collective state can no longer be expressed as a simple function of their individual states. If we have $N$ many qubits, then after we applied a series of gates, our collective qubit state is a linear combination of $2^N$ many distinct states. <u>Measurement</u> is a process that collapses the linear combination of the $2^N$ states into a single one state among them[4]. The single state present after the collapse is the output state of the computation. So, while we operate on exponentially many states at once in a sense when we apply gates throughout the circuit, we can only output a single state in the end. Worse yet, the output state is completely randomly chosen according to a probability distribution that depends on the coefficients in the linear combination of the $2^N$ states. So, if the output is not useful, then one needs to run the algorithm again. Our selection of gates decides the probability distribution on the coefficients, but the gates have certain limitations on them as well. The design of quantum algorithms is now the art of applying gates — chosen from a typically pretty small, limited set of possible gates — in a way so that the probability of getting a useful output state when measuring is sufficiently high. This inherent limitation of quantum computing — together with the common assumption that quantum computers will remain slower and more expensive than classical computers — hints at an important observation: Quantum computers are not general-purpose super computers, rather they are potential special-purpose machines for certain problems where we can leverage their peculiar nature through clever quantum algorithms. Shor's quantum algorithm in Section 3.2 is precisely such a clever algorithm that solves two central problems whose intractability public-key cryptography relies on for its security. Another potential very important application of sufficiently powerful quantum computers is physics simulations, an original intended application of quantum computing was indeed to simulate quantum mechanics itself [138]. Aaronson suggests that common misconceptions about the power of quantum computers originate from focusing only being able to operate on "exponentially many states at once", and then jumping to the incorrect conclusion that quantum computers will be a kind of superior universal super computers [63].

There are several reasons why the circuit model is an idealized model of quantum computation

> The qubits in the circuit model, hereafter referred to as <u>logical qubits</u>, are assumed to be perfect. In a real quantum computer, we would use some sort of real <u>physical qubits</u> to model a typically much smaller number of logical qubits in the circuit model. Physical qubits are unstable to some degree and gate operations can introduce errors. Thus, we need to interleave our higher layer logical operations in the circuit model with lower layer error correction to the physical qubits to ensure that our logical qubits remain uncorrupted. Kalai has conjectured that the error rate in the computations will be a fundamental roadblock that cannot be overcome to actually achieve large-scale quantum computers [21]. Galbraith comments that, in general, theoretical objections to quantum computing do not seem to be widely believed [84]. See Figure 2 for an illustration of the different layers that are envisioned to be part of a potential CRQC.

---

[4] The previously described phenomena of entanglement can now be seen to the right in Figure 1: if we would measure the first qubit and see that it is in state $\bar{0}$, then we would also know that the second qubit is in the same state $\bar{0}$. The corresponding holds if we measure and see that the first qubit is in the state $\bar{1}$. As we measure the first qubit, the whole collective qubit state collapses to a state that is consistent with the measurement result in the first qubit.

➢ In the circuit model we can apply a gate, say a gate that operates on two qubits, to any individual two qubits regardless of their placement vertically in the circuit. In a real quantum computer, interaction between (physical) qubits through gates is sometimes limited to qubits which are physically close to each other in the underlying quantum hardware [40][45][5].

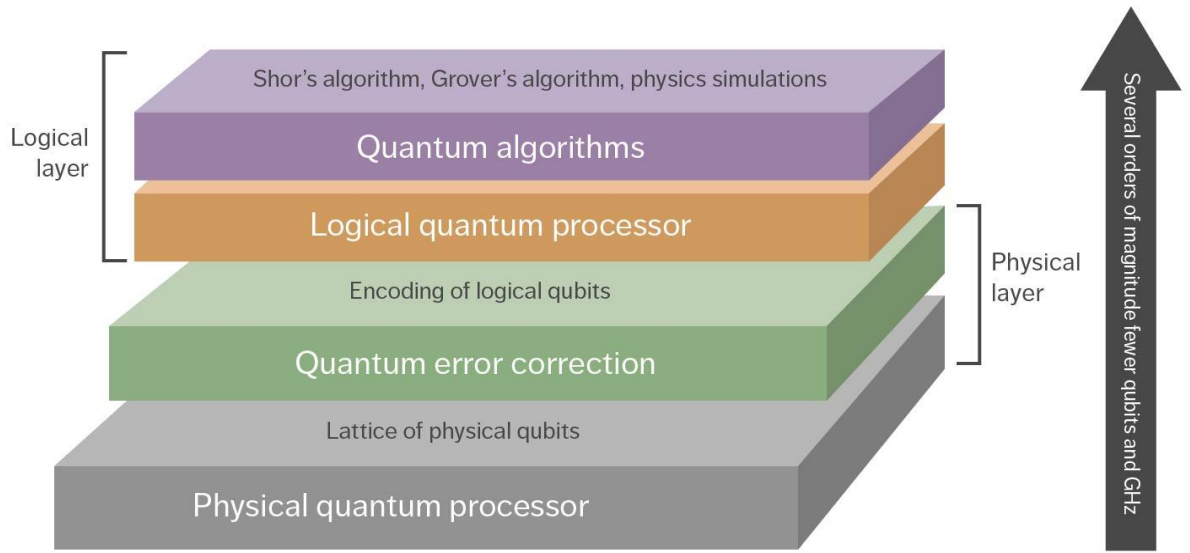

Figure 2: Envisioned structure of future error-corrected quantum computers.

Assuming that the first issue can be overcome, these issues still add overhead to the running time of our quantum algorithms. This is especially relevant to an algorithm such as Grover's algorithm which has an extremely long running time to begin with for cryptographic problems of interest.

The running time of the quantum algorithm is the <u>depth</u> in the number of gates of the circuit (so the depth is 2 in Figure 1). The gates may be applied in parallel if they involve distinct qubits. We do not include the overhead discussed in the previous two bullet points in the depth. That overhead will be dependent on the implementation of the quantum computer and will be added on top of our (logical) depth. The <u>width</u> of the circuit is the number of logical qubits (so the width is 2 in Figure 1). However, the actual number of physical qubits, that in a sense is the true width, will again depend on the implementation of the quantum computer and the error correction implementation, and is an additional unknown overhead. As mentioned in Section 3.2, a CRQC with thousands of logical qubits is often estimated to consist of millions of physical qubits. The error correction may be done in layers where each layer uses error correction on qubits of poorer quality to provide qubits of higher quality to the layer above [129]. There can be a tradeoff between what qubit error rate a specific error correction code can tolerate and the overhead in terms of for example the number of physical qubits necessary to implement the code.

---

[5] To solve this, we can swap the position of two qubits (in the circuit) using a few applications of the quantum "CNOT" gate [48]. CNOT is essentially a quantum version of the XOR operation, and we can use the classic XOR swap trick [47].

For details on the quantum computational model, the lecture notes by de Wolf [85] are a good, up-to-date, and concise starting point.

In the next two subsections we will survey the two quantum algorithms that affect cryptography the most. One thing to beware of is research discussing quantum attacks that rely on having access to a *quantum* implementation of a cryptographic algorithm keyed with a target secret key that the attacker attempts to recover. This threat model appears to be of limited interest since our cryptography will continue to run on classical computers for any foreseeable future.

## 3.2    Shor's quantum algorithm

> ➢ If CRQCs are built, then Shor's quantum algorithm will break today's public-key cryptography. Variants of the algorithm apply to both the discrete logarithm problem (which breaks elliptic curve cryptography and finite field Diffie-Hellman) and to factoring large integers (which breaks RSA).
> ➢ Shor's algorithm is efficient (relatively shallow depth in the circuit model). In the case of using Shor's algorithm to attack RSA, it is essentially as efficient as performing an RSA signing/decryption operation, but on a quantum computer.

Shor's algorithm is certainly the most important quantum algorithm with regard to its impact on cryptography. Shor's algorithm is also one of the most important quantum algorithms in general, since it was this algorithm that showed in 1994 that large-scale quantum computers can efficiently solve important problems (such as factoring very large integers) which appear to be intractable on classical computers. Shor's algorithm consists of a quantum subroutine and some post-processing that can be done on a classical computer. The quantum subroutine for the variant of the algorithm which factors integers is shown in Figure 3. The most expensive operation in the circuit of Figure 3 is the sequence of gates on the bottom wire(s), the ones whose names start with "U". This sequence of gates performs a modular exponentiation modulo $N$, where $N$ is the number (e.g., RSA modulus) that we are trying to factor. Just as for classical computers, this can be done efficiently[6] using e.g., the square-and-multiply algorithm. This means that Shor's algorithm is essentially as efficient as performing an RSA signing/decryption operation, but on a quantum computer. This is typically estimated to take a few hours on a CRQC [130]. However, the width of the circuit

---

[6] E.g., in time complexity that grows slower than $\log^3 N$.

(number of wires) is thousands of qubits[7] for sizes of $N$ that are relevant to cryptography. As discussed in Section 3.1 such qubits are typically envisioned to be implemented through error correction on layers of lower layer qubits (physical qubits in Section 3.1) of poorer quality. By typical estimates we then expect to need millions of such lower quality qubits in the lower layers (which are not visible in Figure 3) [129][20][100][128]. The rightmost boxes in Figure 3 are measurement operations, as discussed in Section 3.1. These operations output classical information from which we have a good probability of being able to factor $N$ after post-processing.

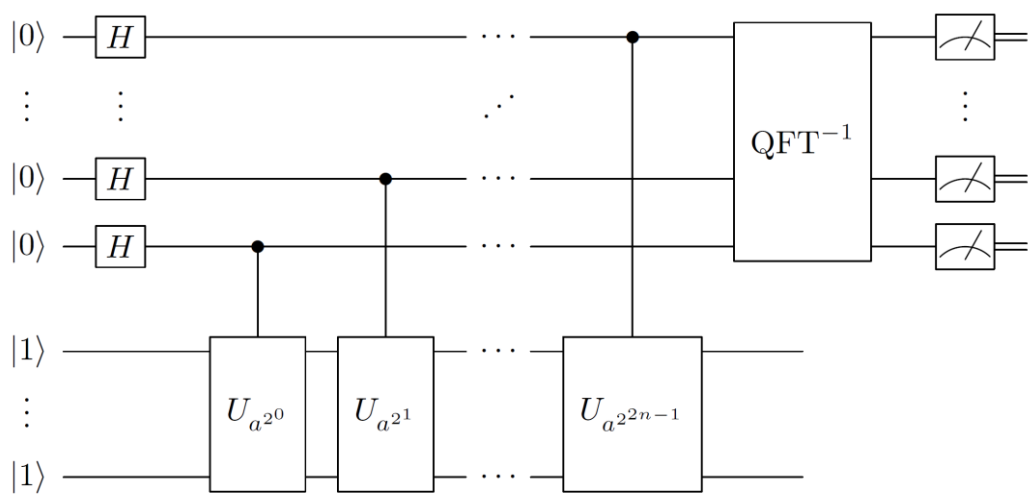

Figure 3: Quantum part of Shor's algorithm (for factoring).

The variant of Shor's algorithm that finds discrete logarithms looks similar but, e.g., in the case of elliptic curve discrete logarithms, we perform two scalar multiplications instead of a modular exponentiation (see Figure 1 in [57]). This results in similar efficiency. The number of qubits required in Shor's algorithm against elliptic curve cryptography (ECC) is estimated to be somewhat lower than the number of qubits required in Shor's algorithm against RSA (factoring) [57]. This is not too surprising since an ECC problem of, for example, size 256 bits is typically estimated to be roughly equivalent in cryptographic strength to an RSA problem of size 3072 bits.This difference in problem size is due to the fact that only very inefficient (exponential) classical attacks are known for ECC [55], while somewhat more efficient (sub-exponential) classical attacks are known against RSA [56]. Thus, we naturally see differences when we also consider quantum attackers that can use Shor's algorithm against both ECC and RSA. However, the number of logical qubits needed for Shor's algorithm against ECC is still estimated to be in the thousands, as for Shor's algorithm against RSA [57].

---

[7] logical ones in the terminology of Section 3.1.

## 3.3 Grover's quantum algorithm and symmetric cryptography

➢ When attacking today's symmetric cryptography, Grover's algorithm is an extremely long-running (extremely large depth in the circuit model) computation that does not parallelize efficiently.

➢ Our most important symmetric primitives, ciphers, MAC algorithms and hash functions are quantum-resistant as is at e.g., the 128-bit security level. In particular, no "doubling of key length" is necessary due to CRQCs.

Grover's algorithm is in a sense an algorithm that finds a needle in a haystack. It quickly puts the collective qubit state in a linear combination of $2^N$ states where each state would be as likely to be outputted if we immediately measured. It then iteratively, roughly $2^{N/2}$ times, applies a set of gates such that each iteration, just so slightly, makes a single target state more likely to be found if we would measure. When we measure in the end, after all $2^{N/2}$ iterations, we find the target state with very high probability. When using this method to attack AES-128, the target state is an encoding of the unknown key which was used to encrypt a target plaintext-ciphertext pair which we assume that we have somehow collected[8]. The previously mentioned "set of gates" used in each of the $2^{N/2}$ iterations are actually an implementation of encrypting the target plaintext and checking if the result matches the target ciphertext. As discussed in Section 3.1 we can do this for each of the $2^N$ states that we start from in parallel, each state being the encoding of a candidate key. Since we have $2^{128}$ candidate keys in AES-128, we have $N = 128$. We then also see that we require $2^{64}$ iterations of this procedure, before measuring in the end. One might ask why we cannot use a set of gates so that we need only 1 iteration instead of $2^{64}$, but this is where the limitations on the set of gates which can be used that we mentioned in Section 3.1 comes in. In fact, one can show that Grover's algorithm solves the generic problem using an optimal number of iterations [60]. The depth of each iteration (measured in the number of serial gates applied) is a problem here. If each iteration contains a serial application of, say $2^{11}$ gates[9], then Grover's algorithm as a whole requires a serial application of $2^{64} \cdot 2^{11} = 2^{75}$ gates. And this is on a single quantum computer, counting only the logical gates in the circuit model. On top of this we have potential overhead from e.g., error correction on the physical qubits as discussed in Section 3.1. In comparison, a classical 5 GHz CPU core executes about $2^{57}$ cycles in a year. Actually, NIST mentions $2^{40}$ as an estimate for the number of serial logical gates that could be applied in a year for presently envisioned quantum computing architectures [59]. These numbers mean that in practice we would need split up the key space and search in each smaller partition of the key space on a separate quantum computer [61].

Grover's algorithm does not parallelize efficiently. For classical computers we can split up a brute force search for a key over $S$ many computers and find the key $S$ times faster. But Grover's algorithm is not a brute force search. In fact, for Grover's algorithm we need $S^2$ quantum computers to find the key $S$ times faster. As an example, under the $2^{75}$ and $2^{40}$ estimates mentioned above, it would take one billion CRQCs, one million years of uninterrupted computation to find a sought AES-128 key. The up-shot is that AES-128 is a relevant security level even in the presence of quantum computers and indeed AES-128 is one of the relevant security categories in the NIST PQC standardization [59]. This means that NIST thinks it is relevant to standardize new public key cryptography that is as costly to break on a quantum computer as it is to do key search for AES-128 using Grover's algorithm.

---

[8] Note that this is a very reasonable assumption. Much information sent encrypted over the Internet is e.g., known headers or plaintext of a known format.

[9] It is roughly an implementation of AES encryption. Gate depth $2^{11}$ for this operation is estimated in Table 8 of [61].

Attacks on MAC algorithms (e.g., HMAC-SHA-256) using Grover's algorithm are implemented similarly to attacks on AES using Grover's algorithm, and the same conclusions hold. This means that our most important symmetric primitives, ciphers, and MAC algorithms are quantum-resistant as is. In particular, no key length changes due to quantum computers are known to be necessary at this point [44]. We think that the view of NIST [44], namely that current deployments using 128-bit symmetric keys do not need to be upgraded because of quantum computers by current knowledge, is well supported by arguments from research as outlined above. Of course, this does not rule out that there could be a quantum algorithm for exploiting a specific weakness in a specific algorithm such as AES-128. But no specific examples of such attacks are known right now for our most relevant symmetric primitives. Newer deployments may want to also consider 256-bit symmetric primitives due to e.g., compliance, and also because the effect on performance may be quite small as commented by e.g., ENISA [64].

### 3.3.1 Grover's quantum algorithm and cryptographic hash functions

➢ There are no known quantum attacks for finding collisions for hash functions that outperform classical attacks for relevant cost measures.

The typical relevant attack against a cryptographic hash function is to find a collision: Two distinct inputs that hash to the same output.

🔍 There is a quite famous variant of Grover's quantum algorithm [87] that finds hash collisions after only roughly $2^{N/3}$ quantum evaluations of the target hash function, where $N$ is the bit-size of the hash digest (e.g., 256 for SHA-256). For an ideal hash function, no classical algorithm can do it in less than approximately an expected $2^{N/2}$ hash function evaluations (the "birthday bound" [88]). This is however only of theoretical interest; the number of hash function evaluations is only indirectly relevant; a directly relevant cost measure is the hardware cost times the running time. As it turns out, the quantum algorithm actually uses roughly $2^{N/3}$ quantum-accessible hardware units as well. So, the hardware cost times running time actually grows roughly like $2^{2N/3}$, and that is on a quantum computer. At the same time, as discussed by Bernstein [89], a classical computing cluster made up of $2^{N/6}$ cheap devices that implement the hash function and work in parallel can find a collision in time proportional to $2^{N/3}$, by together performing the total $2^{N/2}$ hash function evaluations[10] that are expected to be necessary by the birthday bound.

---

[10] The bitcoin network computes approximately $2^{92}$ SHA-256 hashes per year [90], well out of reach of the expected $2^{128}$ needed to find a collision this way for SHA-256. Note that, for example for SHA-1, cryptanalytic progress specific to SHA-1

The fact that quantum attacks against hash functions are no better than classical attacks is visible in the NIST PQC standardization security levels. As explained in Section 4.5, these levels correspond to the optimal attacks (quantum or classical) for key recovery against an ideal block cipher (exemplified by AES) or collision finding for an ideal hash function (exemplified by SHA-2). The attack costs for SHA-2 are based on the most efficient classical attack; no quantum attacks are mentioned.

## 3.4 Building quantum computers

- ➤ It is very unclear when, or even if, a CRQC will ever be built. The gap between today's quantum computers and envisioned CRQCs is huge, and the field faces some near-term challenges such as for example no known applications for the Noisy Intermediate-Scale Quantum (NISQ) computers that are expected to be built these coming years.
- ➤ The best current estimate we have is that an expert committee in 2019 concluded that the emergence of a CRQC during the next decade would be highly unexpected.

The collective effort of working towards quantum computers that can execute large-scale circuits in the circuit model is broad and complex. There are several publicly known engagements both in academia and industry [93][92]. There is also a variety of potential quantum computer realizations (e.g., in terms of how to realize the physical qubits and quantum gates) being studied and proposed, with superconducting qubits and qubits based on trapped ions being popular candidates. However, there is a huge gap between today's noisy small quantum computers and envisioned CRQCs [26]. IBM wrote

> *"Knowing the way forward doesn't remove the obstacles; we face some of the biggest challenges in the history of technological progress."*

when presenting a roadmap for scaling quantum technology in 2020 [99]. The latest and best estimate we have is that a committee of experts from academia and industry concluded in a 2019 report that the emergence during the next decade of a CRQC is highly unexpected [26]. The same report states that there are no known practical applications for the Noisy Intermediate-Scale Quantum (NISQ) computers that we may see in these coming years. Since the report argues that it was the virtuous cycle of incomes and reinvestments that allowed transistor counts in classical computers to grow according to Moore's law for decades, a key finding of the report is

> *"Research and development into practical commercial applications of noisy intermediate-scale quantum (NISQ) computers is an issue of immediate urgency for the field."*

The report argues that the field of building quantum computers may become dependent strictly on government funding otherwise, if useful applications in the near-term cannot be found. Aaronson said in 2021 [137] that there is...

---

allowed researchers to craft a collision in $2^{63}$ hash function evaluations [91], which is better than the $2^{80}$ generic bound described above.

*"Some hope to eke out a near-term advantage [over classical computers] for e.g., quantum simulation with little or no error correction. But claims that we know how to get near-term speedups for optimization, machine learning, etc. are >95% BS!"*

As explained in Section 3.1, quantum computers are not general-purpose super computers, rather they are potential special purpose machines for certain problems where we can leverage their peculiar nature through clever quantum algorithms. Aaronson thinks that the original intended application of quantum computers – simulating quantum physics – will be the "killer app" for quantum computers [2].

Besides technical and usability arguments, judging the progress of quantum computers is further complicated by other aspects. Pornin discusses how judging the progress of quantum computers involves in fact the people – who not only have the best insight into the progress – but are also the same ones who depend on funding to keep the ball that is this very expensive research endeavor rolling [95]. Essentially, what actors in the security community (i.e., industry, academia, governments, and standardization organizations) are doing is watching each other to assess how other actors appear to judge the risk. We can then note that the security community as a whole appears to be calmly awaiting the outcome of NIST PQC standardization that we will discuss in the next section.

Media reporting on quantum computers naturally often focuses on simple metrics such as physical qubit count. As we have discussed, qubit count is not the only relevant metric, but we note that current quantum computers typically have about 100 physical qubits [99][100][135]. We talked about in Section 3.2 how a CRQC is estimated to require a number of physical qubits in the millions. Given the short history of quantum computers and the many competing technologies, it is difficult to estimate how qubit counts will grow the coming years and decades [98]. Simply assuming a Moore's law type scaling in physical qubit counts, it would take 25-30 years to go from 100 qubits to millions of qubits. Of course, there could also be for example improvements in quantum error correction techniques. IBM's roadmap in 2020 aimed to unveil a 127-qubit machine by 2021, a 433-qubit quantum computer by 2022, and a 1121-qubit quantum computer by 2023 [99]. The 127-qubit machine was recently unveiled in 2021 as planned in the roadmap [135].

A claim that a team of researchers had shown "Quantum supremacy" through a computation on a quantum computer consisting of 54 physical qubits caused some stir in 2019 [18]. Quantum supremacy is the achievement of doing a computation on a quantum computer that cannot be achieved in any feasible time on a classical computer. The computation does not have to be useful for anything else than showing Quantum supremacy and the computational task can apparently be chosen in a biased way so that it is naturally much easier for the specific quantum computer at hand than for classical computers[11]. Borcherds criticized this notion of Quantum supremacy by arguing that Quantum supremacy failed at the teapot test [133]. Borcherds explained that teapots achieve "Teapot supremacy" over classical computers because it is very hard to compute on a classical computer exactly how a teapot will shatter if it is tossed in the floor. However, using a teapot, we can compute this by tossing the teapot in the floor and observing the result. While teapots do achieve Teapot supremacy, they are clearly not superior computing devices compared to classical computers. Since Quantum supremacy is similarly biased in the favor of quantum computers, the concept of Quantum supremacy fails in Borcherds's teapot test. Aaronson on the other hand, suggested that we should feel some excitement about the Quantum supremacy experiment [19], seeing it as an important milestone for the field. Aaronson does however stress that the gap between the Quantum supremacy machine and CRQCs is huge and that the research community has no idea how long it will take to close that gap [20]. Some researchers believe that the gap might never be closed in practice [21][22]. Dyakonov [22] says:

> *"I believe that, appearances to the contrary, the quantum computing fervor is nearing its end. That's because a few decades is the maximum lifetime of any big bubble in technology or science. After a certain period, too many unfulfilled promises have been made, and anyone who has been following the topic starts to get annoyed by further announcements of impending breakthroughs. What's more, by that time all the tenured faculty positions in the field are already occupied. The proponents have grown older and less zealous, while the younger generation seeks something completely new and more likely to succeed."*

---

[11] In this case the computational task was to estimate the probability distribution that repeatedly executing a short sequence of quantum gates on a certain number of physical qubits in a specific quantum computer architecture and then measuring the result gives rise to. The classical hardness of the problem appears to still be debated, see [127].

As mentioned above, somewhat similar concerns are raised in an article [25] drawing from statements in the expert committee report [26] on the progress of quantum computing:

> *"The kind of machine that might soon be built, something the committee calls a "noisy intermediate-scale quantum computer," or NISQ computer, probably isn't going to be of much practical use. "There are at present no known algorithms/applications that could make effective use of this class of machine," says the committee. That might change. Or it might not. And if it doesn't, it seems unlikely that industry will keep investing in quantum computing long enough for the technology to pay dividends."*

In their FAQ on Quantum Computing and Post-Quantum Cryptography, NSA states [104]

> *"NSA does not know when or even if a quantum computer of sufficient size and power to exploit public key cryptography … will exist."*

It is important to keep in mind that NSA, in one of its roles, provide guidance and recommendations for protecting US National Security Systems (NSS). Information in these systems that is encrypted under public-key cryptography today may require confidentiality for many decades. The potential damage CRQCs may inflict here is the core of the motivation to seek for counter measures even if we have uncertainties around when and if these computers can be built. The risk that CRQCs will be built means that currently deployed public-key cryptography needs to be replaced with quantum-resistant alternatives.

For further reading on the state of building quantum computers, the summary in the expert committee report [26] (and the whole report) is highly recommended.

# 4     Post-Quantum Cryptography (PQC)

> ➢ The NIST PQC standardization is an effort to standardize new quantum-resistant public-key cryptography. In particular, these algorithms can execute completely in software on classical computers.

In this report we define Post-Quantum Cryptography (PQC) as quantum-resistant cryptography that can serve as a functional replacement for current public-key cryptography. The NIST PQC standardization is an effort to standardize new Key Encapsulation Mechanisms (KEM) and digital signature algorithms that fall within the PQC scope [96]. In particular, these algorithms can execute completely in software on classical computers, in contrast to e.g., Quantum Key Distribution (QKD) which requires custom hardware. A signature algorithm is an established primitive, but the KEM abstraction shown in Figure 4 has not been used as extensively in the past.

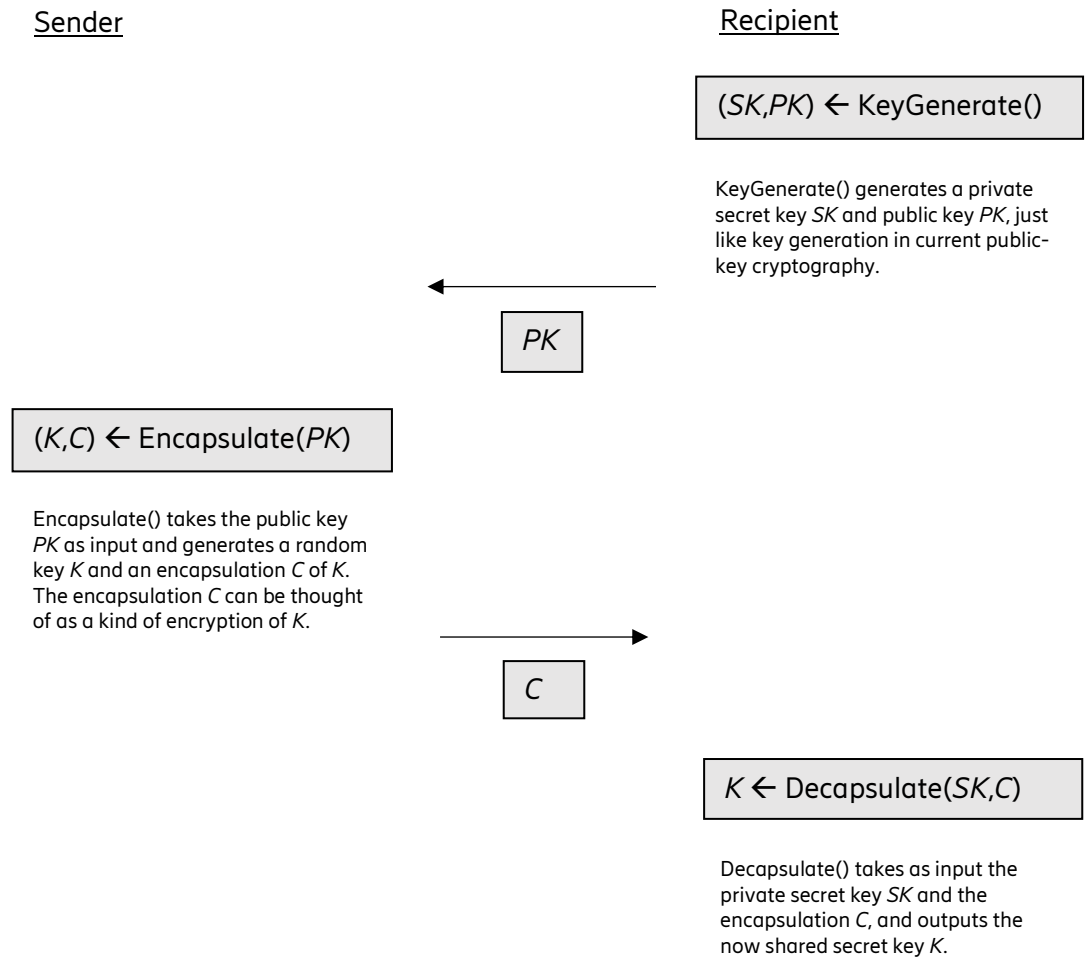

Sender                                    Recipient

$(SK,PK) \leftarrow$ KeyGenerate()

KeyGenerate() generates a private
secret key $SK$ and public key $PK$, just
like key generation in current public-
key cryptography.

$PK$

$(K,C) \leftarrow$ Encapsulate($PK$)

Encapsulate() takes the public key
$PK$ as input and generates a random
key $K$ and an encapsulation $C$ of $K$.
The encapsulation $C$ can be thought
of as a kind of encryption of $K$.

$C$

$K \leftarrow$ Decapsulate($SK,C$)

Decapsulate() takes as input the
private secret key $SK$ and the
encapsulation $C$, and outputs the
now shared secret key $K$.

Figure 4: The operations and communication of a KEM. Sender and recipient share a secret key $K$ after
carrying out the KEM procedure. The key $K$ can then be used to e.g., key a symmetric cipher.

Note from Figure 4 that, for example, both

> ➢ key establishment through RSA (where the sender would encrypt a randomly chosen key $K$ under the RSA public key of the recipient and send that as the encapsulation $C$ to the recipient), and
> ➢ key exchange through Diffie-Hellman (DH) (where the sender would send an ephemeral public key $C$ as the encapsulation to the recipient from which the recipient can compute the key $K$ as the shared secret using the DH primitive) fit within the KEM abstraction. We note one important distinction, in the DH example above, the encapsulation may also be a static public key of the sender — effectively mapping a Non-Interactive Key Exchange (NIKE) [46] to the KEM setting. NIKE, which is used in for example Group OSCORE [121], is one example of how the KEM and signature frameworks targeted by the NIST PQC standardization do not fill all the needs where public-key cryptography (especially elliptic curve cryptography) is used such as for example Privacy-Enhancing Cryptography [119], Identity-Based Encryption and Signature Aggregation [120]. The KEMs in the NIST PQC standardization provide KEMs that are secure both in a stronger security model that allows for using a static recipient public key (so-called IND-CCA(2) security[12]) and KEMs with less overhead that support only certain ephemeral usage.

## 4.1    Why the wait?

> ➢ Establishing trust in new cryptographic primitives takes time. Especially for public-key cryptography where typically a new class of mathematical problems must be analyzed and well understood to ensure security.
> ➢ Elliptic curve cryptography is too efficient, flexible, and rich to give up easily.
> ➢ Updating standards and libraries takes time, and updating procedures and protocols is costly.
> ➢ Hybrid solutions can offer important additional assurance when migrating to PQC.

One might ask why actors are waiting to implement PQC if the threat of quantum computers cannot be ruled out and if information that is confidentiality protected today through public-key cryptography can be recorded and potentially attacked in the future. The main reasons are

---

[12] This gives security against not only passive attackers, but also active attackers. To achieve such strong security, the randomness used in creating the encapsulation is revealed to the recipient during the decapsulation procedure. The recipient then asserts that the encapsulation was generated in an honest manner using this randomness. This is the so-called Fujisaki-Okamoto transform [97] which is a generic method to strengthen the security of the recipient's static key by making sure that only honestly generated encapsulations are completely processed during decapsulation (see also [122] for variants). In contrast, for DH we have instantiations such as X25519 [51] where it is secure to reuse a key pair in multiple DH key exchanges without using a transform like Fujisaki-Okamoto. These DH instantiations are the result of carefully developing parameter sets and (domain-aware) checks that avoid and mitigate known attacks, see for example [66]. This can naturally be more efficient than a generic mitigation through a transform like Fujisaki-Okamoto. Renes suggests that the heavy usage of the Fujisaki-Okamoto transform can make the NIST PQC candidates more vulnerable to certain physical side-channels such as power analysis [116].

➢ Establishing trust in cryptographic primitives is a slow process. Ideally, we want as many experts as possible to try to break a primitive for as long time as possible before we put it into production use. Some of the PQC systems are relatively new and for new public-key cryptography there is typically a new class of mathematical problems that must be analyzed and well understood to ensure security. After the security community jointly establishes trust in a system, actors also need to wait for standardization. Part of the trust-establishment step has merged with standardization in the case of the NIST PQC standardization. After standardization, actors additionally need access to well-reviewed libraries and hardware that can support operations with the new cryptographic algorithms. Standardization also may take place in stages; first the cryptographic primitives may be standardized, then protocols such as TLS may be updated to support them.

➢ There is a cost associated with the PQC systems. Elliptic curve cryptography (ECC) provides a wide variety of very efficient primitives. We also have very efficient and secure implementations and standards available. Nobody wants to give up ECC[13]. As we will see, one typical difference between PQC and ECC is larger keys or signatures. Furthermore, changing algorithms (affecting public keys and certificates), and updating security critical software and hardware obviously costs money. There is thus a balance between prudent preparations for switching to quantum-resistant cryptography on the one hand, and making sure that the investment in implementing quantum-resistant cryptography will be a long-term secure and good choice, on the other hand.

🔍 A potential solution to the issue in the first bullet point is to run two cryptographic systems in parallel, one that is quantum-resistant and one that is well-studied and trusted to be secure against classical adversaries. So, for example, we could derive a final session key in a key establishment protocol from a combination of $K_{PQC}$ and $K_{ECDH}$, where $K_{PQC}$ is the key from a PQC key establishment mechanism and $K_{ECDH}$ is the shared secret from an elliptic curve Diffie-Hellman key exchange. This is for example what was done when Cloudflare and Google tested PQC schemes in TLS [102], and NIST plans to incorporate this possibility into a future revision of SP 800-56C that deals with key establishment [101]. An attacker would need to break both primitives independently to recover the final derived key. This is sometimes called a hybrid scheme, but note that hybrid is used for other things as well in cryptography. The obvious downsides to the hybrid solution are poorer efficiency, additional communication overhead, and additional code complexity. These downsides can be manageable though and hybrid solutions can offer important additional assurance when migrating to PQC. One could consider a similar hybrid solution for signatures as well [1][101], however, here the overhead would be more significant if hybrid signatures were used throughout a long certificate chain. Also, incorporating PQC into encryption is more urgent since encrypted information typically needs confidentiality protection for a longer time than a public key is valid for signature verification. At the same time, updating deployed systems for especially PKI can take many years.

---

[13] Lattice-based cryptography, on which some of the most important PQC proposals are based, also seems to be very flexible and rich, exemplified by the introduction of the first fully homomorphic encryption system in 2004 by Gentry, which was based on lattice cryptography [109].

An alternative quantum-resistant solution that could be deployed today already is including Pre-Shared (symmetric) Keys (PSKs) in session key derivation [65] for communication protocols such as TLS. The serious downsides to this solution are the need to distribute pair-wise secret PSKs to any two parties that want to communicate, and the need for an endpoint to store many confidential PSKs rather than a single private key for public-key cryptography.

## 4.2 Fitting PQC into existing protocols

> The main difference with fitting PQC into existing protocols is larger public key or signature sizes. The performance of the PQC schemes with regard to running time is often comparable to that of today's public-key cryptography.

Taking TLS as an example, a KEM could replace the key establishment done through Diffie-Hellman key exchange today in TLS 1.3. For forward secrecy[14], the client could send an ephemeral KEM public key to the server to which the server responds with an encapsulation of a shared a key $K$. The session key is then derived from $K$, much like it is derived from the shared Diffie-Hellman secret in TLS 1.3 today.

In TLS, the handshake is authenticated by the server towards the client using a digital signature over all the information exchanged during the handshake. The client authenticates towards the server either through digital signature, or in the application layer (by e.g., "logging in") as is often the case in human-to-server communication on the Web. These signatures could of course be replaced by PQC signatures. But for this kind of pair-wise (interactive) communication there is also another possibility. The server could authenticate by proving knowledge of a key $K$ that is decapsulated from an encapsulation which is encapsulated towards a KEM public key tied through PKI to the server. This means that we replace a signature sent from the server with an encapsulation sent from the client to the server and e.g., a MAC tag computed using $K$ (over the information in the handshake) sent from the server to the client. Of course, the most relevant overhead in practice may be a certificate chain that certifies the server public key rather than the size of the signature or KEM public key used by the server for authentication. See [33] for a discussion about using KEMs for handshake authentication in TLS. Existing versions of TLS could also have relied on authentication through static DH keys (instead of RSA-based signatures), but one problem with this would be the overwhelming popularity of certificates with RSA keys. Static DH keys are used for authentication in for example Signal X3DH [123] and EDHOC [124]. Note that the DH primitive is more flexible than a KEM. In the X3DH and EDHOC examples, an ephemeral DH public key (which can be thought of as the encapsulation when mapping to a KEM) can be combined with multiple different DH keys to form shared secrets, something which is not possible with KEM encapsulations in general. As applications migrate to PQC, new certificates will be needed anyway and maybe there will be more widespread usage of KEMs for pair-wise authentication.

---

[14] Forward secrecy typically implies that new ephemeral public keys are used for key establishment in essentially every session. After the session ends, all keys are securely deleted. Thus, session keys cannot leak at a later time. Such ephemeral public keys must of course be authenticated by some means to provide any security.

As shown in Section 4.5 the main difference with fitting PQC into existing protocols is communication overhead. The performance of the PQC schemes with regard to running time is typically comparable to that of today's public-key cryptography. NIST summarizes the status of replacing current public-key cryptography with any of the different (classes of) PQC algorithms as

> *"Unfortunately, each class has at least one requirement for secure implementation that makes drop-in replacement unsuitable."* [67]

The earlier mentioned PQC TLS experiment by Google and Cloudflare gives some observations about using PQC in TLS [118][16]. In this experiment, Diffie-Hellman key exchange in TLS 1.3 was experimentally replaced with two hybrid alternatives: a) ECDH + NTRU-HRSS-KEM[15], and b) ECDH + SIKE. As can be seen in Table 1 of Section 4.5, alternative b) has smaller communication overhead but slower running times, while alternative a) has the opposite properties. While the clients were biased towards more powerful platforms (x64 and AArch64) — which could favor the slower running times of alternative b) — the overall TLS handshake times tended to be faster for alternative a).

## 4.3    Fitting PQC into constrained scenarios

Over constrained network links, the typically larger signature and encapsulation sizes of the PQC schemes may lead to for example packet fragmentation and increased energy spent in the radio or wire-line interfaces on sending more bits.

For constrained devices, there has been quite a lot of implementation work on the 32-bit Cortex M4. In the NIST PQC standardization third round conference, the pqm4 project reported on the performance of implementations of almost all the finalist candidates in the standardization [114]. The results where for a platform with the Cortex M4, 128 KB RAM and 1 MB of flash storage. On the KEM side, SABER performed well, while the results on the signature side were more mixed. Implementations can be challenged by the fact that the RAM is not even large enough to hold the whole public key for some of the standardization candidates. In [115], implementers work with Classic McEliece (whose public key is 260 KB for its smallest parameter set) on a platform with 192 KB of RAM. In the call for proposals, NIST asks implementers to also consider even more constrained 16- and 8-bit microprocessors [113]. Atkins [112] says that RAM appears to be the bottleneck with storage currently growing at a faster rate over time. So, Atkins suggests that NIST should focus on RAM usage in the standardization, rather than code size, when considering devices that are even more constrained than typical platforms with the Cortex M4.

---

[15] Now merged into the finalist candidate NTRU in Section 4.5.

## 4.4    Stateful hash-based signatures

> ➢ Stateful hash-based signature schemes depend on security-critical dynamic state. Otherwise, their security depends only on some generally accepted assumptions about cryptographic hash functions.
> ➢ The standardized stateful hash-based signature schemes are currently the only option for applications that
>    o can handle the dynamic state,
>    o cannot wait for the outcome of the NIST PQC standardization, and
>    o are prepared to deal with using a stateful algorithm that may not yet be supported by common tools for PKI.

Two stateful hash-based signature schemes have been standardized by IETF and NIST[16]: LMS and XMSS. We discussed in Section 3 how the security of hash functions is unaffected by quantum computers. These schemes are indeed PQC schemes, and this is the reason why there was a sudden interest to standardize them. The idea of using hash functions to build signature schemes is otherwise quite old but has not been interesting up to now since signature sizes are much larger than for signature schemes currently deployed and the stateful hash-based signature schemes have the serious burden of depending on security critical dynamic state. We explain the idea behind the standardized stateful hash-based signature schemes in Appendix B.

It is critical for security that two distinct messages are not signed using the same private key state (leaf WOTS private key in Appendix B). This dynamic security-critical state is the reason that NIST and NSA consider these schemes not to be suitable for general use [41][42]. These schemes, LMS and XMSS, are currently the only option for applications that can handle the dynamic state, cannot wait for the outcome of the NIST PQC standardization and are prepared to deal with using a signature algorithm that may not be supported by tools for PKI.

## 4.5    NIST PQC standardization — candidates and progress

> ➢ Lattice-based cryptography will offer a good middle-way for PQC with efficient running times and medium-sized communication overhead. These schemes are represented both as finalist KEM and signature algorithm candidates in the standardization.

NIST aims to announce the first selected algorithms for standardization close to the end of 2021 and provide draft standards in 2022-2023[17]. From the initial 69 accepted 1st round candidates, the 3rd round candidates consist of 7 finalists and 8 alternate candidates. The first algorithms selected by NIST will mainly be chosen from the finalists. Additional algorithms may be selected after an additional round of evaluation. Four of the finalists are KEMs and three are signature schemes. NIST has successfully attracted world-leading experts in the field of PQC to make submissions in the standardization.

---

[16] See [110] for a comparison between LMS and XMSS written by authors from the LMS team.

[17] Unless otherwise stated, the implicit reference in this section is the NIST Status Update on the 3rd Round [69] from the Third PQC Standardization Conference (June 7-9, 2021).

To assess the security of the PQC candidates, NIST has asked the submissions to rank the security of the candidates according to security levels that are based on the optimal classical and quantum attacks on ideal ciphers (exemplified by AES) and hash functions (exemplified by SHA-2). For example, the schemes in Table 1 and 2 are believed to be security level 1. This means that any significant quantum or classical attack on them should be as expensive as a key recovery attack on AES-128 (using either classical brute-force or Grover's quantum algorithm) [59]. Some experts have argued that security level 1 or 2 should not be used due to cryptanalysis for the PQC candidates still not being completely settled down [136].

> Three of the finalist KEMs (Kyber, NTRU, SABER) are so-called structured lattice-based schemes. NIST expects to select at most one of these KEMs for standardization, but may standardize more than one. Lattice-based cryptography uses hard problems on mathematical objects called lattices as the foundation for its security [71]. While security reductions that relate the security of some of these lattice-based schemes to known conjectured-to-be-hard computational problems on lattices exist, for efficiency, concrete schemes are typically instantiated with heuristic assumptions and parameters are set only with respect to best known attacks in practice [72]. An example of a heuristic assumption could be that a certain structure in a problem does not give an attacker any advantage compared to the corresponding unstructured mathematical problem. Overall, there is quite strong confidence in the security of lattice-based schemes. Cryptographic applications have been studied extensively the last two decades [71] and an original NTRU system dates back to the 90's [73]. NSA recently stated that it expects that lattice-based cryptography as standardized by NIST will be deemed secure enough to protect US National Security Systems (NSS) [42]. In fact, NSA plans to add structured lattice-based cryptography to its CNSA suite already at the end of the third round of the NIST PQC standardization (tentatively late 2021 – early 2022) [104]. It appears that structured lattice-based cryptography will offer a good middle-way for PQC with efficient running times and medium-sized communication overhead. As can be seen in Table 1, the public keys and encapsulations in these finalist KEMs start in the 600–800-byte range. These quantities are 32-64 bytes in today's elliptic curve Diffie-Hellman, which is vulnerable to Shor's quantum algorithm.

> The fourth finalist KEM – Classical McEliece – is code-based. In code-based cryptography, error-correcting codes are used to decode a message that has been masked with random noise. Classical McEliece uses more unstructured codes than some other code-based candidates that have not qualified as finalists, and is based on a system by McEliece from 1978 [74]. The system enjoys a long and stable security history [75], which NIST says inspire confidence [76]. Classical McEliece is one of two PQC systems that the German BSI have issued an initial recommendation for[18]. As can be seen in Table 1, Classical McEliece has small encapsulation size (128 bytes), but very large public keys at roughly 200 KB. NIST says [77] that it was impressed by Post-quantum WireGuard where static Classical McEliece keys are used long-term for pair-wise authentication and their cost is amortized over many handshakes. This is the same principle as discussed in Section 4.2 of using KEMs for pair-wise authentication.

---

[18] The other one also being a conservative choice, FrodoKEM – an unstructured lattice-based scheme that is also a NIST PQC 3rd round alternate candidate.

➢ A noteworthy alternate candidate KEM is SIKE (Supersingular Isogeny Key Encapsulation). SIKE works on elliptic curves like elliptic curve Diffie-Hellman (ECDH) but instead of staying on a single fixed curve, SIKE jumps between many different curves that are connected through so-called isogenies. In particular, unlike ECDH, SIKE does not rely on the hardness of the elliptic curve discrete logarithm problem which is broken through Shor's quantum algorithm. SIKE is based on problems that are relatively new, but it has a very stable (albeit short) security history [78]. As can be seen in Table 1, SIKE has relatively small public keys and ciphertexts (roughly 400 bytes, and they can be compressed down to roughly 200 bytes), but orders of magnitude slower running time than many other candidates.

| KEM algorithm | Generate key | Encaps. | Decaps. | Public key size | Encaps. size |
|---|---|---|---|---|---|
| NTRU (ntruhps2048509) | 0.048 ms | 0.0073 ms | 0.012 ms | 699 B | 699 B |
| Kyber (kyber512) | 0.0070 ms | 0.011 ms | 0.0084 ms | 800 B | 768 B |
| SABER (lightsaber2) | 0.012 ms | 0.016 ms | 0.016 ms | 672 B | 736 B |
| Classic McEliece (mceliece348864) | 14 ms | 0.011 ms | 0.036 ms | 261120 B | 128 B |
| SIKE (SIKEp434_compressed) | 3.0 ms | 4.4 ms | 3.3 ms | 197 B | 236 B |
| ECDH (X25519) (non-PQC) | 0.038 ms | 0.044 ms | 0.044 ms | 32 B | 32 B |
| ECDH (P-256) (non-PQC) | 0.074 ms | 0.18 ms | 0.18 ms | 32-64 B | 32-64 B |
| RSA-3072 (non-PQC) | 400 ms | 0.027 ms | 2.6 ms | 384 B | 384 B |

Table 1: Single-core median performance on Intel Xeon E-2124 3.3 GHz of some of the NIST PQC KEM algorithm candidates (and some current non-PQC alternatives) at NIST PQC security level 1. Source: r24000, supercop-20210604 at [107]. The measurements for RSA-3072 are estimated by taking the measurements for verify and sign from Table 1 as encapsulate and decapsulate, respectively. The measurements for SIKE are for a different platform: Intel Core i7-6700 3.4 GHz [108].

As stated earlier, there are three different finalists in the digital signature track. NIST is concerned about the lack of diversity in these finalists (two being structured lattice-based schemes and the third having an unstable security history), and has therefore declared that there will be a new call for proposals for additional quantum-resistant digital signature schemes at the end of the third round of the original standardization effort. The third round is expected to end in late 2021 or early 2022.

➢ Two of the finalist signature schemes Dilithium and Falcon are structured lattice-based schemes. They vary in their construction, and Falcon has significantly smaller signatures (roughly 800 bytes). In their talk at the Third PQC Standardization Conference, the Dilithium team pointed out that the design choice of Falcon that enables small signatures also makes secure implementation of the scheme significantly more complicated and difficult than in the Dilithium case [80]. Like for their KEM counterparts, there is quite strong confidence in the security of these structured lattice-based signature schemes, and they have had a stable security history through the NIST PQC standardization. Like for the structured lattice-based KEM schemes, NIST expects to standardize at most one of these two schemes and that scheme has a good chance to be the main general purpose PQC signature scheme considered for adoption in protocols and industry in the years following the NIST PQC standardization.

➤ The third finalist signature scheme is Rainbow which is based on multivariate cryptography. Rainbow's security relies on the hardness of finding solutions to certain polynomial equations with certain structure. There have been relevant attacks on Rainbow during the standardization process [81]. The Rainbow team responded to the latest attack by only updating the way that attacker cost is estimated in their security model, thus leaving their parameter sets unchanged [82]. NIST has expressed concern about the security of multivariate cryptography [83], and ENISA states that "the security analysis of Rainbow cannot be considered stable at the moment" [81]. As can be seen in Table 2, Rainbow has large public keys (roughly 160 KB), but very small signatures (64 bytes — as small as current elliptic curve cryptography signatures).

➤ SPHINCS+ is an alternate candidate and a stateless hash-based signature scheme. It is based on the same idea as the stateful hash-based signatures of Section 4.4. As explained in Appendix B, the security critical state in stateful hash-based signatures controls that internal low-level one-time signature keys are used only once. In SPHINCS+ these one-time keys are replaced by "few-time" signature keys, and for each signing operation in SPHINCS+, such an internal few-time signature key is randomly chosen. Each few-times signature key can be reused up to a certain limit while retaining security of the overall algorithm. Overall, the algorithm supports for up to $2^{64}$ signatures to be issued, as requested by NIST in the call for proposals [96]. As can be seen in Table 1, SPHINCS+ has a slow signing operation and large signatures.

Detailed documentation for all of the NIST PQC standardization candidates is available from the NIST third round home page [111]. Being written by experts in the field and continuously updated throughout the standardization effort that started in 2017, this documentation is a great technical entry point to the field of PQC. Many submissions also include optimized constant-time[19] implementations.

---

[19] A constant-time implementation is one where the running time of the algorithm does not depend on the value of secret key data. This prevents timing-based side-channel attacks.

| Signature algorithm | Generate key | Sign | Verify | Public key size | Signature size |
|---|---|---|---|---|---|
| Falcon (falcon512dyn) | 5.9 ms | 0.23 ms | 0.029 ms | 897 B | 666 B |
| Dilithium (dilithium2aes) | 0.015 ms | 0.041 ms | 0.019 ms | 1312 B | 2420 B |
| Rainbow (rainbow1aclassic363232) | 2.7 ms | 0.017 ms | 0.0087 ms | 161600 B | 64 B |
| SPHINCS+ (SPHINCS+-SHA-256-128s-simple) | 27 ms | 210 ms | 0.28 ms | 32 B | 7856 B |
| LMS (using SHA-256, limited to $2^{20}$ messages) | - | - | - | 56 B | 2828 B |
| Ed25519 (non-PQC) | 0.014 ms | 0.015 ms | 0.050 ms | 32 B | 64 B |
| ECDSA (P-256) (non-PQC) | 0.029 ms | 0.041 ms | 0.086 ms | 64 B | 64 B |
| RSA-3072 (non-PQC) | 400 ms | 2.6 ms | 0.027 ms | 384 B | 384 B |

Table 2: Single-core median performance on Intel Xeon E-2124 3.3 GHz of some of NIST PQC signature algorithm candidates (and some current non-PQC alternatives) at NIST PQC security level 1 (Dilithium is at that submission's smallest suggested parameter set — at level 2). Source: r24000, supercop-20210604 at [103]. The measurements for SPHINCS are for a separate platform: Intel XeonE3-1220 3.1 GHz in Table 6 of [106]. LMS is a stateful hash-based signature scheme (see Section 4.4), these schemes have slow key generation, while signing and verification takes at most a few milliseconds on a comparable platform to those used by the other algorithms in the table.

## 4.6 How to prepare for PQC

The security community and relevant actors are waiting for the NIST PQC standardization to conclude. For example, NSA still recommends a suite of non-quantum-resistant public-key algorithms for protecting TOP SECRET material [43]. Such material typically needs to be protected for decades. What industry can do today is to ensure that products are sufficiently prepared for migrating to public-key algorithms with properties (key and signature/ciphertext sizes) similar to those for the round three proposals in the NIST PQC standardization when the time comes. The schemes in the NIST PQC standardization are intended to be drop-in replacements with regard to functional interfaces for the current public-key key establishment and signature algorithms. This means that protocols such as IKEv2 and TLS can continue to work as they do today with slightly different performance characteristics in the asymmetric (public-key) part of the protocols when the protocols and certificates are updated to support the new algorithms. We mentioned in Section 4.1 how Google and Cloudflare have run experiments over the Internet to try and determine the impact of deploying PQC KEMs in TLS. One consideration for that experiment is whether maximum transmission unit restrictions over the network in combination with communication overhead from PQC algorithms give rise to additional packet fragmentation during the TLS handshake [16]. The communication overhead of PQC can also contribute to filling congestion windows in protocols such as TCP [134].

Exactly what will happen when the NIST PQC standardization ends sometimes in the next few years is hard to say. A likely scenario is that other standards-developing organizations will follow and that important protocols such as TLS and IKEv2 will be updated to support the new standardized PQC algorithms. It is also likely that the new algorithms have good library support in important software libraries such as OpenSSL at that time since work on production grade (e.g., efficient, and secure against timing side-channel attacks) implementations is already in progress[28]. In contrast, somewhat more uncertain is how fast deployment in applications, support in hardware and support in PKI will come. These are things that come at a cost (e.g., overhead, issues with legacy interworking, and deployment/development costs) to the relevant actors, and one important drive here could be what real progress is made towards large-scale quantum computers in the next few years. As a rough analogy one can consider how slow applications and actors have historically been at phasing out algorithms (e.g., MD5, RC4, SHA-1) whose security has gradually but publicly degraded through cryptanalysis [15]. However, one must not only consider the cost of upgrading cryptographic algorithms and the security risk of using degraded algorithms, but also the pure cost in reputation of using degraded algorithms. An event that could drive forward PQC adoption after the NIST PQC standardization is complete is if SDOs and other important actors such as NSA not only update standards and guidance to support the new PQC algorithms, but also deprecate our currently used public-key algorithms. NSA says [104]

> *"Programs should anticipate that after NIST provides the needed [PQC] standards there would be rapid movement toward requiring support of a quantum-resistant standard in new acquisitions."*

As discussed previously in this report, updating authentication is typically less pressing than confidentiality protection, and thus PKIs might be relatively slow in deploying PQC, as briefly discussed by Langley [118]. Firmware updates with very long-term keys is a case of authentication that requires special attention. Stateful hash-based signatures — being standardized already — have been considered for this use case [105].

We cannot rule out that results in cryptanalysis for some class of the new PQC algorithms could also appear, such an event could seriously disrupt deployment. Most serious would presumably a serious attack on the lattice-based schemes be since these are represented through several contributions in the NIST PQC standardization and are typically regarded to be a good middle way when taking communication overhead, performance, and security into account, as discussed in Section 4.5. It should also be noted that for most schemes in the NIST PQC standardization, the underlying computational assumptions have been studied for many years, if not decades, before the standardization started.

---

[28] This can be seen e.g., from the activities discussed in the NIST PQC standardization forum [14].

A NIST whitepaper Getting Ready for Post-Quantum Cryptography [67] recommends that enterprises, as an initial step in migrating from current public-key cryptography to PQC, makes an inventory of where and for what they are using public-key cryptography. After that, [67] advises that enterprises determine use characteristics such as hardware/software limits related to current key sizes, latency, and throughput thresholds, etc. NIST is also launching a project to ease migration to PQC [68].

## 4.7 Possible adoption of quantum-resistant cryptography in future 3GPP networks

The cryptographic protection in the radio access of 3GPP networks relies mostly on symmetric cryptography today (e.g., UE authentication and radio access ciphering/integrity protection). As we saw in Section 3.3 there is no need to update this protection specifically due to the threat of large-scale quantum computers[21]. In contrast, the SUPI protection (SUCI [125]) that relies on public-key cryptography is affected by quantum computers and so is any 3GPP network domain security that relies on public-key cryptography (e.g., IKEv2 in the IPsec suite and TLS) as explained in the previous section. A possible upgrade path for SUCI is HPKE [126] (a public-key encryption system based on the KEM+DEM paradigm which is under development in the IETF) which is expected to be updated with PQC algorithms at the same time as TLS.

## 5 Summary of the security of Quantum Key Distribution (QKD)

- ➢ QKD typically needs to rely on many of the same computational assumptions as conventional efficient cryptography.
- ➢ QKD has less well-understood of a risk profile and implementation security than conventional cryptography.
- ➢ QKD is inherently tied to custom hardware. This can increase cost and risk, compared to conventional cryptography that can often be patched and upgraded in software.
- ➢ QKD is inherently tied to the physical layer, this gives a quite different attack surface than conventional cryptography.
- ➢ QKD is fundamentally a point-to-point protocol. This can imply a dependency on trusted intermediate nodes that is not compatible with the modern technology environment which is moving towards end-to-end encryption and Zero Trust principles.
- ➢ QKD may be more sensitive to Denial-of-Service attacks than conventional cryptography.
- ➢ There is a consensus in the security community that QKD has many fundamental issues that would need to be solved before being considered as a secure complement to conventional cryptography.

QKD is the primary example of so-called Quantum Cryptography but it is important to understand that it has nothing necessarily to do with quantum computers. The idea was first described in the 80's [34] and several commercial systems exist. We refer the interested reader to Wikipedia for a simple description of QKD using photons [34]. Essentially, in theory, QKD allows Alice and Bob to agree on a shared (classical) key by Alice sending qubits (photons in the Wikipedia example) to Bob over a quantum communication channel. By sending and receiving the photons using a clever encoding scheme Alice and Bob can determine with almost certainty if an attacker has interfered and eavesdropped (i.e., measured the qubits). If this is the case, they abort. Otherwise, they can derive a short key that a potential attacker has supposedly essentially no information about. An important condition is that Alice and Bob need an *authenticated* conventional communication channel (in addition to the quantum one) to make comparisons and exchange information. Also, harmless noise on the quantum channel and potential interference by an eavesdropper are indistinguishable.

---

[21] However, if user authentication is augmented with current public-key cryptography to provide additional security even in very strong threat models, then these parts would need to be updated.

🔍 It is sometimes said that QKD is provably secure from "the laws of physics" or similar (see e.g., [23] for references). Bernstein [23] criticizes these kinds of formulations as misleading. Indeed, if the security of QKD is provable from the laws of physics, then it sounds like QKD is a well-defined physical process that has some proven physical security property. But the published physical attacks on commercial QKD (see e.g., [24] for references) contradict this. Bernstein in fact argues that it is unlikely that a physical process similar to how QKD is typically described could be absolutely provably secure since information about the agreed upon (classical) shared key presumably leaks (to some small, albeit non-zero amount) through various standard physical phenomena such as e.g., electromagnetic radiation[22]. The way that we will make sense of the connection between the security proofs for QKD and the physical systems claiming to perform QKD in the rest of this report is the following. There is a protocol called QKD taking place in a mathematical model that is modelled after quantum mechanics. Attackers are limited in the mathematical model to certain well-defined mathematical actions and physical leakage through side-channels or other physical attacks on the protocol is out of scope in the model. Presumably, the theoretical model cannot be implemented absolutely securely in the physical world due to e.g., side-channel attacks, as argued by Bernstein. Then we have hardware components provided by actors that are claimed to provide a sufficiently secure implementation (a kind of practically realizable approximation) of the theoretical protocol[23]. As we will argue below, this is where part of the security problems with QKD begin since the most important attack surface against modern cryptography is implementation details.

QKD is sometimes said to be provably secure without relying on computational assumptions. At the same time, we mentioned above that QKD needs external authentication. For this reason, QKD has been described as a "key expansion primitive" [9]: A short pre-shared symmetric key used for authentication is expanded through QKD into an arbitrarily long shared secret key[24]. The argument here is then that the resulting shared secret key is everlasting secure, it cannot be cracked in the future in contrast to e.g., a PQC KEM key that could be cracked if someone recorded the KEM protocol messages and in the future discovered a break-through efficient algorithm to solve the underlying mathematical problem. The same argument of everlasting security afterwards has been put forth when e.g., a PQC signature algorithm is used to authenticate QKD [9], instead of a shared symmetric key. As discussed in the previous paragraph, it is unlikely that any physical realization of QKD actually provides truly absolute "everlasting security"[25]. But let us leave this aside for a moment and discuss the relevance of the fact that QKD does not rely on a computational assumption the same way that a PQC KEM does. It is important to understand that relying on an "extra" computational assumption (that the PQC KEM will not be completely broken during the protection lifetime of the data) is only one out of many concerns in choosing a cryptography suite. As an example, current public-key signature schemes depend, not only on the hardness of some mathematical problem such as factoring integers, but also on cryptographic hash functions. At the same time, we have stateful hash-based signature schemes with well-understood theoretical security which rely only on hash functions for their security. You could argue that using stateful hash-based signature schemes and relying only on hash functions would be more daunting to adversaries since it relies on fewer computational assumptions. Despite this, hash-based signature schemes have seen essentially zero use due to the drawbacks discussed in Section 4.4, and there is little library support. Many attackers would presumably rather attack a fragile proprietary implementation of a signature scheme depending on security-critical dynamic state, rather than joining the long line of mathematicians that have tried to solve e.g., factoring large integers efficiently. Furthermore, any complex system must presumably rely on many computational assumptions in very many places anyway (storage encryption, firmware updates, etc.).

---

[22] See also the assumption about "Sealed laboratories" in device independent QKD later in the text.

[23] It is common to have security proofs (in a mathematical model) building on computational assumptions for conventional cryptography as well. In the gap, between that security proof and a concrete implementation executing in a given threat model, lives side-channel attacks that depend on implementation details which are out of scope in the mathematical model.

There is a consensus in the security community that QKD has many fundamental issues that would need to be solved before being considered as a secure complement to conventional cryptography [30][28][31][27][12][29]. As already hinted at in the discussion above, the security-relevant criticism against QKD includes:

➢ QKD needs external authentication as explained above. This either implies relying on computational assumptions such as PQC signature schemes or using more impractical symmetric key distribution schemes. Depending on this choice the system as a whole may no longer be "unconditionally secure". As we argued above, it is unlikely that the theoretical protocol can be mapped to a physical process in an absolutely secure way anyway.

➢ Current QKD systems typically rely on ordinary symmetric cryptography for cryptographic high-speed data protection. Such symmetric cryptography is not unconditionally secure, which – once again – means that the system as a whole is not unconditionally secure.

---

[24] It is sometimes further envisioned that part of the agreed upon shared secret key can be used to authenticate the next QKD protocol run [9]. This means that, in theory, we no longer need to rely on the original pre-shared symmetric key after using it once.

[25] See also [23] where Bernstein discusses an attacker that records physical side-channel information about the processing of the agreed upon shared key and then derives the shared key by computations on the side-channel information.

➢ While the security proof for QKD is a good first step – just like such a tool is valuable for conventional cryptography – attacks on modern cryptography almost always target implementation details, and not the underlying theoretical algorithms. The long and rich history of attacks on implementations of cryptographic protocols show how hard it is to get all of these implementation details exactly right. The standard cost-efficient way to reduce the risk of implementation vulnerabilities is to use well-reviewed and highly trusted implementations of cryptographic primitives such as those in a well-known open-source software library. Regarding the implementation security of QKD hardware, NSA [12] states that

> *"Communications needs and security requirements physically conflict in the use of QKD/QC, and the engineering required to balance these fundamental issues has extremely low tolerance for error. Thus, security of QKD ... is highly implementation-dependent rather than assured by laws of physics."*

Indeed, there have been attacks on commercial implementations of QKD [24]. It is important to note that conventional cryptography is implemented at a higher layer than the physical layer. In modern cryptography, the physical layer is often just an untrusted pipe and the attacker may be assumed to be able to launch any attack against it. In contrast, QKD is inherently tied to the physical layer and its practical security depends on implementation details at the physical layer as shown by attacks on commercial QKD systems [37]. This fact gives QKD and conventional cryptography quite different attack surfaces in practice. Experimental "device independent" QKD (e.g., DI-QKD, DDI-QKD and MDI-QKD) is being studied by the QKD community, it appears that this technology does not close this attack surface against QKD [49] since these experimental technologies still need to make various assumptions about isolation of the quantum devices:

> *"In device-independent quantum key distribution, we make the additional assumption that there is no communication between the adversary and the quantum devices."* [38]

> *"In terms of security, MDI-QKD closes all side-channels in the detection unit, which significantly simplifies the path towards achieving implementation security in QKD, as now one only needs to secure the source. MDI-QKD requires, however, that certain assumptions on the sources are satisfied."* [39]

➢ Post-Quantum Cryptography (PQC) typically has a better understood risk profile than QKD [12]. Much of the knowledge about implementing public-key cryptography securely which has been gradually (and sometimes painfully) built over the last decades can be more or less directly transferred to the PQC schemes in the NIST PQC standardization, this includes security models and low-level implementation details such as protection against side-channel attacks. Furthermore, while PQC relies on the intractability of certain mathematical problems, this still appears to be a better situation than QKD which is inherently tied to hardware. A mathematical problem can be made available to, and discussed over the Internet between any researchers that are interested while every piece of QKD hardware is a unique physical thing that may require expensive hardware to thoroughly study.

➢ <u>QKD is inherently tied to custom hardware.</u> This is in contrast to PQC systems in software that can be patched or upgraded in software. This difference may negatively affect the security of QKD, leaving vulnerabilities deployed longer in production because of replacement costs. Both conventional cryptography and QKD may depend on hardware acceleration for e.g., high-speed data protection through symmetric cryptography.

➢ <u>QKD is fundamentally a point-to-point protocol.</u> The straightforward way to extend it is to use a sequence of QKD protocol instances between trusted nodes that form a path between the two end-user parties. Introducing trusted nodes can be a serious obstacle in a technological environment where many applications are moving more towards end-to-end security [35] and Zero Trust principles. Experimental ideas for building "quantum repeaters" that supposedly remove the need for trusted nodes are also being considered [36]. It is critical that any such technology is securely implemented, and not only modelled after something that is secure in some idealized mathematical model. In contrast, conventional cryptography can secure information that is sent through untrusted nodes over the Internet today already. Figure 5 shows a comparison between secure two-party communication using conventional crypto (e.g., PQC) and using QKD.

➢ <u>QKD may be more sensitive to DoS (Denial-of-Service) attacks than conventional cryptography [12].</u> NSA [12] specifically mentions the fact that sensitivity to an eavesdropper is at the core of the security idea for QKD. At the same time, in principle, any communication system (including classical ones) can be subjected to DoS if the attacker has sufficient control over the channel. Classical communication can however be error-corrected through standard ways to mitigate attacks to a certain degree. It may also be easier to utilize redundant paths for classical communication which is more flexible than QKD communication which is tied to hardware and particular communication channels.

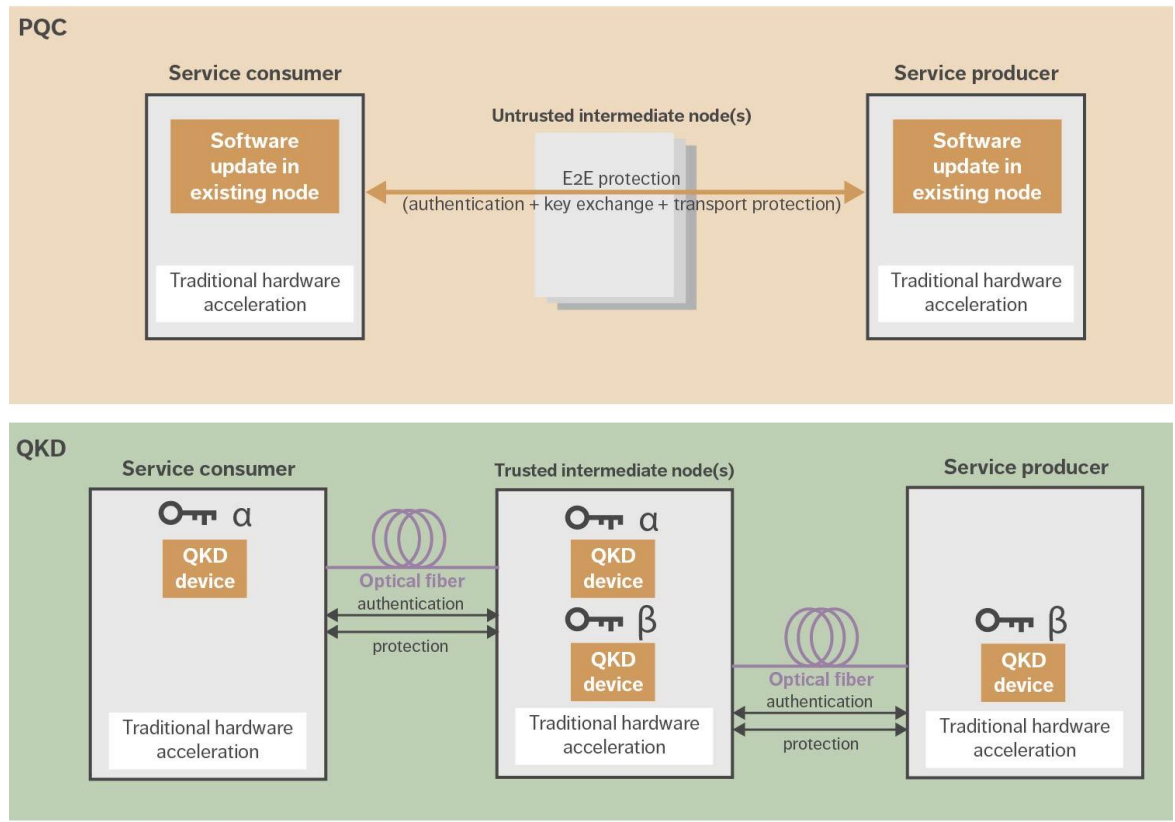

Figure 5: Secure communication comparison between conventional cryptography (e.g., PQC) and QKD. The orange boxes in the PQC scenario show how PQC could be introduced into a legacy environment that uses today's public-key cryptography.

It remains to be seen whether the strong reservations against QKD put forth by government agencies will alone be a show-stopper to deploying QKD in any critical national infrastructure. In [12], NSA writes that it does not recommend deploying QKD for securing US National Security Systems, unless the limitations discussed in [12] are overcome. NCSC (UK) states [27]:

> "NCSC does not endorse the use of QKD for any government or military applications, and cautions against sole reliance on QKD for business-critical networks, especially in Critical National Infrastructure sectors."

ANSSI (France) highlights a number of problems and limitations of QKD but still does not want to rule out that QKD could be deployed as a defense-in-depth measure on point-to-point links in combination with classical cryptography (e.g., a classical asymmetric key exchange such as elliptic curve Diffie-Hellman) as long as *"the cost incurred should not jeopardize the fight against current threats to information systems"* [32]. However, if one is actually seriously concerned about relying on a computational assumption used in a specific instance of a public-key cryptography scheme, then a more cost-efficient solution is likely to deploy several public-key primitives in a hybrid fashion as discussed in Section 4.1. For key establishment, an extremely paranoid application could bundle up say RSA encryption, DH,

ECDH and all of the round 3 NIST PQC KEM contributions in a hybrid solution so that an attacker would need to break all of the schemes to gain any useful information at all about the shared key. It is hard to see a competitive business application in any industry deploying such a paranoid solution when most of the daily TLS traffic over the Internet rely on a single version of some Diffie-Hellman key exchange.

# 6      Appendix A: Quantum random number generators (QRNGs)

A hardware Random Number Generator (RNG) is random bit generator that uses some physical process to collect random bits. A quantum RNG (QRNG) is an RNG that specifically uses a physical process that is at the "quantum level" [8].

➢ Quantum random number generators solve no real issue with our current hardware RNGs. If trustworthy vendors make QRNG technology in the future that is as well-understood and certified to the same degree as common RNG technology, then QRNGs could be evaluated as alternatives to common RNG technology.

🔍 The claimed argument for preferring QRNGs over other more traditional hardware RNGs is that QRNGs in theory produce perfectly random bits while the physical process utilized in traditional RNGs may produce biased bits that are not perfectly random. However, this is in practice a non-issue since we have established methods for going from biased non-perfectly random bits to what we need: bits that look perfectly random in the eyes of any existing adversary. To give more detail: the physical process in an RNG need not produce perfect random bits, rather it must produce a relatively small amount of *total* randomness (e.g., 128-256 bits of so-called entropy) so that we can smooth it out and output as much cryptographically secure randomness we need through cryptographically secure pseudo-random number generators (CSPRNGs) (which are seeded and reseeded using random bits from the physical process in the RNG). The principles that the CSPRNG process builds on are typically the same ones as our conventional symmetric cryptography builds on.

One issue that has been discussed about traditional hardware RNGs is trust in the non-maliciousness of the RNG [4]. As we argued in Section 5 regarding QKD, relying on an opaque piece of hardware for security operations is typically less appealing than trusting an open-source software library. But in the case of RNGs or QRNGs, we have no choice, we must trust an opaque piece of hardware in practice[26]. In practice this trust falls on the manufacturer of that hardware. If we ignore test procedures that check if the RNG is not faulty, this is essentially a trust issue that is not solved by technology. On their recent CPU chips, Intel provides a hardware RNG that is accessible through the RDRAND instruction. This traditional RNG follow the principles from the previous paragraph [7] and most discussion about it has been focused on whether one wants to trust that this RNG is not malicious [4]. One can hedge against non-ideal trust in a single RNG vendor (e.g., Intel) by mixing output entropy from more than one hardware source in some well-reviewed way[27]. Similarly, one can and should mix in accessible randomness from system events that are believed to contain some entropy to avoid relying only on a single hardware source of entropy. It should also be noted that security typically relies on trusting at least some part of the hardware, e.g., the CPU. Building secure complex systems at a competitive price will imply trusting many other components, including various software libraries (e.g., compilers, programming language libraries, operating system distributions, etc.). For an industry product that has already decided to trust for example the CPUs from a vendor, relying also on the RNGs on those CPU chips may be reasonable and cost-efficient. Trust in a hardware vendor here means trust in that the supply chain and production (including delivery) of the product is not malicious. If we also consider vulnerabilities inadvertently inserted in the random generator, then it appears that one should also prefer a well-known RNG such as the Intel one, rather than one from some smaller less established QRNG vendor.

---

[26] Even if the design of the piece of hardware is published, how do you as a business verify (in a commercially competitive way) that the instance of the hardware that you're holding is completely and accurately described by the design?

[27] While there may still be theoretical attacks on this in a very strong threat model where a malicious randomness source can (in real-time) spy on the output of all other sources [5], this is the best idea we have of how to solve this very hard problem.

Additionally, a main general practical problem is to make proper use of hardware random bit generators in virtual and container implementations. See the work of BSI [117], which discusses amongst other the impact of the VMM (Virtual Machine Manager) on the randomness in the virtual machine.

NCSC lists a better understanding of the robustness and security of QRNG technology as a future research challenge [50]. On the topic of QRNGs, the NSA points out that there is a variety of non-quantum hardware RNGs that have been appropriately validated and certified [104].

To conclude, QRNGs solve no real issue with our current hardware RNGs. If trustworthy vendors make QRNG technology in the future that is as well-understood and certified to the same degree as common RNG technology, then QRNGs could be evaluated as alternatives to common RNG technology.

# 7 Appendix B:  Idea behind the standardized hash-based signature schemes

The basic underlying primitive is called a Winternitz one-time signature (WOTS) and its private key is for one-time use. The idea is roughly that we reveal pre-images for the hash function such that iteratively hashing a pre-image a certain number of times we obtain a corresponding digest in the public key. The number of times we iterate the hash function is decided by the corresponding nibble[28] value in the digest of the message being signed. Since finding pre-images for hash functions is infeasible, no one can forge signatures. This means that if we sign message digests that are 32 bytes (e.g., as for SHA-256) which is 64 nibbles, then the one-time public key consists of about 64 digests. The size of 64 number of 32-byte digests is 2048 bytes. But a single one-time signature is not very useful on its own. Instead, one uses a hash tree (often called Merkle trees) whose leaves are WOTS public keys. The public key of the signature scheme is the root of the hash tree and a signature consist roughly of

(i)     a WOTS signature related to a leaf one-time public key, and
(ii)    $L$ digests which are needed to verify the leaf one-time public key's place in the hash tree which have $L$ levels.

---

[28] A nibble is a sequence of four bits.

A candidate WOTS public key is computed from (i) (and the message being verified) as described above, and that candidate WOTS public key is then verified by using (ii) and the candidate WOTS public key to compute a candidate root value of the hash tree. The signature is valid if the candidate root value equals the public key (of the signature scheme). We are skipping some details that are important for security here. Furthermore, we may actually have multiple layers of trees so that we do not need to compute all the bottom leaf nodes at key generation, this choice would depend on how many messages we intend to sign over the lifetime of the public key. If we have, say 20 levels, in the hash tree ($L = 20$), then we can sign $2^{20}$ messages and a signature has size $2048 + 20 \cdot 32 \approx 2700$ bytes.  In comparison, an ECDSA signature is typically 64 bytes. If we have two layers of trees, then signatures are twice as big, and we can sign $2^{40}$ messages instead. See Table 2 for exact public key and signature size for the LMS signature scheme using a single tree and 20 levels. That instance can use its private key to sign at most $2^{20}$ messages. Stateful schemes like LMS and XMSS are out of scope in the NIST PQC standardization.